# Manipulating Momentum-Space and Real-Space Topological States in Metallic Strontium Ruthenate Ultrathin Films

Xuan Zheng[1,2,3], Zengxing Lu[1,2], Bin Lao[1,2], Sheng Li[1,2], Run-Wei Li[1,2,3*], Milan Radovic[4*] and Zhiming Wang[1,2,3*]

[1]CAS Key Laboratory of Magnetic Materials and Devices, Ningbo Institute of Materials Technology and Engineering, Chinese Academy of Sciences, Ningbo 315201, China
[2]Zhejiang Province Key Laboratory of Magnetic Materials and Application Technology, Ningbo Institute of Materials Technology and Engineering, Chinese Academy of Sciences, Ningbo 315201, China
[3]Center of Materials Science and Optoelectronics Engineering, University of Chinese Academy of Sciences, Beijing 100049, China
[4]PSI Center for Photon Science, Paul Scherrer Institut, CH-5232 Villigen PSI, Switzerland
*Email: zhiming.wang@nimte.ac.cn, milan.radovic@psi.ch, runweili@nimte.ac.cn

**Abstract**

SrRuO$_3$, a 4d transition metal oxide, has gained significant interest due to its topological states in both momentum space (Weyl points) and real space (skyrmions). However, probing topological states in ultrathin SrRuO$_3$ faces challenges such as the metal-insulator transition and questioned existence of skyrmions due to possible superposition of opposite anomalous Hall effect (AHE) signals. To address these issues, we investigate ultrathin SrRuO$_3$/SrIrO$_3$ heterostructures and their AHE and topological Hall effect (THE). Our results reveal metallized ultrathin SrRuO$_3$ down to the monolayer limit with an AHE signal. ARPES measurements confirm the metallic and topological band structure of ultrathin SrRuO$_3$. Furthermore, the AHE sign remains negative over a wide thickness range, where THE is still observed. This observation excludes the two-channel explanation of THE and provides evidence for the existence of skyrmions in ultrathin SrRuO$_3$.

# Introduction

Topological states of matter have emerged as a vibrant research frontier in contemporary condensed matter physics and material science [1,2]. Such topological phases encompass electronic band structures supporting exotic quasiparticles in momentum space, as well as real-space topological spin textures and magnetic skyrmions toward spintronic devices [2–6]. As a 4d transition metal oxide, the ferromagnetic metal $SrRuO_3$ is attracting substantial renewed interest as a host for realizing both momentum-space topological semimetals and real-space magnetic skyrmionic phases [7,8]. Theoretical works have proposed $SrRuO_3$ could harbor topological semimetal states characterized by Weyl points and Dirac nodal lines near the Fermi level [7,9]. Recent experimental transport and ARPES measurements have also revealed signatures consistent with magnetic Weyl semimetals in $SrRuO_3$ thin films [10–16], adding support to the predicted topological electronic structure. The topological electronic states contribute large Berry curvature, which manifest as a highly tunable anomalous Hall effect (AHE), which is sensitive to magnetic order and external parameters [9,17–20]. Such non-monotonic changes of the transport properties have been interpreted as signatures of magnetic Weyl nodes in momentum space [7,9]. Additionally, $SrRuO_3$ possesses strong spin-orbit interactions that, when coupled with interfacial symmetry breaking, induces sizable Dzyaloshinskii-Moriya interactions which facilitate stabilization of magnetic skyrmions. The rich interplay of lattice structure, SOC, magnetic order, and electron correlation in $SrRuO_3$ makes it an ideal platform to explore both topological electronic states and magnetic textures, such as skyrmions, which have been recently observed via the topological Hall effect (THE) in $SrRuO_3$ thin films and heterostructures [8,21–28].

Despite these intriguing observations, the microscopic origin of the THE in $SrRuO_3$ remains a topic of intense debate [29]. Some studies suggest that the measured THE signals reflect the intrinsic detection of skyrmions [8,21–25], while others propose that the signals arise from an inhomogeneous anomalous Hall effect due to effects like thickness variations, defects, and interface modification in the films [19,30–39]. Addressing this controversy is crucial both for critically evaluating the microscopic mechanisms producing the topological Hall signatures, as well as unambiguously demonstrating and stabilizing skyrmionic spin textures for next-generation spintronic devices. However, detailed examinations elucidating the intrinsic versus extrinsic contributions pose significant materials challenges. While few unit-cell SrRuO$_3$ films present alluring opportunities for both device integration of skyrmions as well as mapping any emergent electronic phases, dimensional tuning induces sharp metal-insulator transitions below 3-4 unit cells that obstruct probing such phenomena [40,41]. Additionally, in the ultrathin limit, the AHE shows a strong dependence on thickness and disorder with possible sign changes. The superposition of opposite AHE signals with different coercive fields could generate hump-like signals resembling a topological Hall effect [32,34]. Using approaches like Berry phase engineering by heterointerface to maintain a fixed AHE sign with thickness could aid in excluding the extrinsic effects. Overall though, the highly tunable AHE makes interpretations complex, with the existence of true topological Hall effect signals still under debate [29]. Achieving metallic conductivity and ferromagnetism in the ultra-thin regime while minimizing AHE variations remains vital for reliable stabilization and detection of skyrmions toward device integration.

To address these challenges and conflicting interpretations, we have designed ultrathin SrRuO$_3$/SrIrO$_3$ heterostructures that maintain metallicity and ferromagnetism in the ultrathin regime while minimizing AHE variations. By combining *in-situ* ARPES and transport measurements, we unveil the preserved topological band structure and the emergence of a robust negative AHE sign over a wide thickness range in ultrathin SrRuO$_3$. Remarkably, we observe hump-like features in the Hall effect without accompanying AHE sign changes, providing compelling evidence for an intrinsic topological Hall effect in our heterostructures. These findings offer new insights into the interplay between topology, magnetism, and electronic correlations in ultrathin oxide systems and pave the way for harnessing their unique properties in future spintronic devices.

**Results**

Ultrathin SrRuO$_3$ films with various thicknesses $n$ are deposited on 20 unit cells (u.c.) of SrIrO$_3$, constructing [SrIrO$_3$]$_{20}$/[SrRuO$_3$]$_n$ (20+$n$) heterostructures as shown in Fig. 1(a). The *in-situ* RHEED oscillation enables control of film thickness with atomic precision as shown in Fig. S1. The temperature dependent resistivity measurement in Fig. 1(b) demonstrates the metallic behaviour of the samples with a visible kink induced by the magnetic phase transition of SrRuO$_3$. Notably, the samples with ultrathin SrRuO$_3$ below 3 u.c. which are insulating without ferromagnetism in single films, also exhibit the kink indicating ferromagnetism. The magnetic measurement of the hysteresis loop in Fig. 1(c) and temperature dependent magnetization in Fig 1(d) further confirms the ferromagnetism in ultrathin $SrRuO_3$ down to monolayer. The Curie temperature ($T_c$) can be estimated from both kinks in the $\rho$-T curve in Fig 1(c) and the M-T curve in Fig 1(d). Fig.S2 and Fig.S3 show the determination of Curie temperature by taking the first derivative in the $\rho$-T curve and two tangent crossings in the M-T curve. $T_c$ ranges from 105-120 K as determined by $\rho$-T curves and spans from 106-130 K in the M-T curves. The approximate correspondence of the $T_c$ determined by the two approaches confirms the kink in the $\rho$-T curve originates from the magnetic phase transition in $SrRuO_3$. This also indicates metallicity of ultrathin $SrRuO_3$ in the heterostructures. Despite previous reports that sub 3 u.c. films undergo metal insulator transition and loss of ferromagnetism [40,41], the mono- and bi-layer $SrRuO_3$ films here are rather metalized with ferromagnetism when interfaced with $SrIrO_3$. The recovery of ferromagnetism and metallicity in these ultrathin $SrRuO_3$ films enables further magnetotransport measurements to probe the topological states, which is infeasible in insulating $SrRuO_3$.

*In-situ* ARPES measurements are employed to directly probe the electronic structure and confirm the metallicity of ultrathin $SrRuO_3$ in $SrIrO_3$/$SrRuO_3$ heterostructures with $SrRuO_3$ thicknesses of 1, 2, 4, and 8 monolayers. As shown in Fig.2, the Fermi surface intensity map varies when the thickness of $SrRuO_3$ changes. By comparing with the Fermi surface maps of a single $SrRuO_3$ layer (Fig. 2(b)) and a single $SrIrO_3$ layer (Fig. S2), the thickness dependent evolution of the Fermi surface map can be attributed to the mixing of signals from the $SrIrO_3$ and $SrRuO_3$ layers for thin $SrRuO_3$ films, and dominant signals from $SrRuO_3$ for thick films. Since the energy dispersion along high symmetric X-M-X direction distinguish single $SrIrO_3$ (Fig. S2(b)) and $SrRuO_3$ (Fig. 2(b)) films, the $SrRuO_3$ band dispersion can be identified in heterostructures of varying thicknesses. This confirms nonzero density of states at the Fermi surface, verifying the metallicity in ultrathin $SrRuO_3$. Comparing the X-M-X energy dispersion maps (Fig. 2(d)), the slight shift of band structure indicates preserved electronic structure down to a monolayer $SrRuO_3$ thickness. The band crossing along X-M-X is actually identified as Weyl points in previous work in collaboration with first-principle density functional theory calculations [12]. Thus the preserved band dispersion along X-M-X suggests the persistence of topological states in ultrathin $SrRuO_3$ and motivate further exploration via transport measurements.

Building upon the confirmed metallicity and ferromagnetism in ultrathin SrRuO$_3$, Hall measurements are performed to investigate the topological states in the system. Since the anomalous Hall effect in SrRuO$_3$ relates to Berry curvature contributed by Weyl points, and the topological Hall effect evidences skyrmions with topological magnetic order, investigating the Hall effect in ultrathin SrRuO$_3$ is crucial for probing its topological states. Hall measurements are systematically performed on ultrathin SrRuO$_3$ with thicknesses of 1,2,3,4,6,8,10 u.c. grown on 20 u.c. SrIrO$_3$, constructing 20+n SIO/SRO heterostructures. The anomalous Hall signal is obtained by subtracting the normal Hall effect. Fig. 2(a) shows the temperature dependent anomalous Hall signal for different heterostructures. Intriguingly, AHE signals can be measured for the first time in mono- and bi-layer SrRuO$_3$. Notably, the AHE sign persists over a wide temperature range for SrRuO$_3$ thicknesses below 8 u.c., in sharp contrast to the single SrRuO$_3$ films and SrRuO$_3$/SrIrO$_3$ heterostructures with ultrathin SrIrO$_3$ below 3 u.c. where only 3 u.c. or 4 u.c. SrRuO$_3$ exhibit AHE with the same sign when changing temperature [8,21,34]. Fig. 3(b) summarizes the temperature dependence of the AHE, showing that the AHE is positive at high temperatures but becomes negative at low temperatures in the 20+10 SIO/SRO heterostructure. The signal change dependent on the temperature is similar to that observed in single SrRuO$_3$ films, indicating the intrinsic mechanism of the observed AHE. Unexpectedly, a persistent negative AHE resistance occurs over a wide range of temperature and thickness range. This preservation of AHE sign may be attributed to interfacial effect of SrIrO$_3$ film and structural modification of thick SrIrO$_3$ film compared to ultrathin SrIrO$_3$ film. The anomalous Hall conductivity is further plotted against magnetization in Fig. 3(c). The conductance is inverted from positive to negative above a threshold magnetization, which is qualitatively consistent with theoretical calculation of Hall conductance contributed by Berry curvature in literature. However, the $\sigma_{AHE} - M$ curves for different heterostructures do not overlap with a trend of shifting to higher magnetization when thickness get thinner. This trend resembles single SrRuO$_3$ film but more gradual with thickness [34], which can be attributed to the interfacial effect of SrIrO$_3$. The interfacial SrIrO$_3$ impedes the sign change of AHE in SrRuO$_3$. As the ARPES result in Fig.2 shows preserved band structure of SrRuO$_3$ down to monolayer, we infer that the thickness dependent band structure in SrIrO$_3$/SrRuO$_3$ heterostructure undergoes gradual shift without abrupt change. The influence of thickness variation and interfacial SrIrO$_3$ on electronic structure is depicted in Fig. 3(d), the interfacial effect of SrIrO$_3$ when reducing thickness triggers band structure shift in opposite to decreasing temperature, different from the case in single SrRuO$_3$ film where an abrupt change in band structure is implied when changing thickness [38]. The observed unconventional AHE behavior and its persistence down to the monolayer limit in SrRuO$_3$/SrIrO$_3$ heterostructures raise intriguing questions about the potential existence of topological Hall effects and skyrmions in this system.

To further investigate the possible presence of topological Hall effects, we examine the AHE loops of 20+6 and 20+8 heterostructures, where a hump-like feature is observed as shown in Fig. 4(a). The hump-like feature is observed for a wide range of temperature from 10 K up to 90 K. Remarkably, the AHE sign persists negative over all measured temperature for these heterostructures (Fig.4(b)), in contrast to the 20+10 sample, which exhibits AHE sign change with temperature and no hump-like feature. These behaviour differs sharply from reported hump observations coinciding with AHE sign changes, which have been attributed to superimposed opposite AHE signals in inhomogeneous films [18,19,33,34,37,42,43]. The persistent negative AHE sign in our heterostructures rules out the possibility of opposite AHE contributions from different thicknesses SrRuO$_3$ layers or from the SrIrO$_3$/SrRuO$_3$ interface. Specifically, the AHE remains negative for SrRuO$_3$ thicknesses n < 9, and even for the 20+10 sample, a negative AHE is observed above 50 K, where no positive contribution from the bulk is expected based on the thickness-dependent AHE evolution (Fig. 3). The exclusion of these extrinsic factors as the origin of the observed hump-like features and persistent AHE sign strengthens the case for an intrinsic topological Hall effect and the existence of skyrmions in our ultrathin SrRuO$_3$/SrIrO$_3$ heterostructures. This finding is particularly significant in the ultrathin limit, where the challenges of metal-insulator transitions and AHE sign changes have previously hindered the unambiguous identification of intrinsic topological Hall effects [21,34,40].

**Discussion**

Our data presented above demonstrate that we have successfully stabilized metallic and ferromagnetic states in ultrathin SrRuO$_3$ through heterostructure engineering with SrIrO$_3$. The metallization of mono- and bi-layer SrRuO$_3$ in our heterostructures contrasts with the metal-insulator transition and loss of ferromagnetism reported in ultrathin SrRuO$_3$ films [40,41], highlighting the critical role of the SrIrO$_3$ interface in stabilizing the electronic and magnetic properties of SrRuO$_3$ in the ultrathin limit. Our *in-situ* ARPES measurements confirm the preservation of the topological band structure in SrRuO$_3$ down to the monolayer limit, which is particularly significant given the sensitivity of topological states to dimensionality and interface effects [44,45]. This finding demonstrates the robustness of topological states in the 2D limit and opens up new possibilities for exploring real- and momentum-space topological phenomena in ultrathin oxide heterostructures. The unique electronic structure at the SrRuO$_3$/SrIrO$_3$ interface, as revealed by our ARPES results, shows a gradual evolution of the band structure with thickness, which helps to maintain a consistent Berry curvature contribution to the AHE.

The observation of a robust negative AHE over a wide thickness range in our SrRuO$_3$/SrIrO$_3$ heterostructures differs markedly from the behaviour reported in single films and other heterostructures, where the AHE sign is often found to change with thickness or temperature [17,23,24,34]. This unusual AHE dependence, combined with the presence of hump-like features in the Hall effect, provides strong evidence for its intrinsic topological nature in our SrRuO$_3$/SrIrO$_3$ heterostructures. The simultaneous observation of hump-like features in the Hall effect and a persistent AHE sign offers crucial insights into the ongoing debate surrounding the intrinsic versus extrinsic origins of the topological Hall effect in SrRuO$_3$ [29,38,39]. Indeed, the persistence of the AHE sign in our heterostructures strongly suggests an intrinsic origin, as extrinsic mechanisms such as the superposition of opposite AHE signals from different layers or domains would be expected to lead to AHE sign changes [32–34]. By carefully controlling the interface structure and electronic properties, we have been able to minimize these extrinsic effects and reveal the intrinsic topological Hall effect. As discussed in recent works [7,9], the topological band structure of SrRuO$_3$ can give rise to a complex Berry curvature landscape, with competing contributions from multiple sources such as Weyl points and nodal lines. The presence of hump-like features in the absence of AHE sign changes suggests that these competing contributions are delicately balanced in our heterostructures, allowing for the manifestation of the topological Hall effect.

**Conclusion**

In conclusion, we have successfully addressed the challenges in probing topological states in ultrathin SrRuO$_3$ layer by constructing SrRuO$_3$/SrIrO$_3$ heterostructures. Interfacing SrRuO$_3$ with SrIrO$_3$ is pivotal in stabilizing the metallic and ferromagnetic states in ultrathin SrRuO$_3$, opening avenues for exploring its topological properties. Our combined experimental approach, encompassing transport measurements, magnetic characterization, and ARPES, provides compelling evidence for the existence of topological states in ultrathin SrRuO$_3$. The preserved metallicity and ferromagnetism of mono- and bi-layer SrRuO$_3$ stabilize its topological band structure. The observation of AHE with a negative sign over a wide thickness range, along with the presence of THE, collectively demonstrate the robustness of topological states in this system. Our findings decisively illuminate the intrinsic nature of the topological Hall effect in ultrathin SrRuO$_3$, establishing a groundbreaking platform to explore the intricate interplay between topology, magnetism, and electronic correlations in the 2D limit. The ability to stabilize and probe topological states in ultrathin oxide heterostructures opens up exciting possibilities for future research and potential applications in spintronics and quantum devices. We anticipate that our work will stimulate further theoretical and experimental investigations into the rich physics of topological phenomena in ultrathin oxide systems and their potential for engineering novel quantum states of matter.

## Methods

**Sample growth.** SrIrO$_3$/SrRuO$_3$ heterostructures were deposited on (001)-oriented SrTiO$_3$ substrate with a pulsed laser deposition (PLD) system using a KrF excimer laser operating at $\lambda = 248$ nm. TiO$_2$-terminated SrTiO$_3$ substrates are obtained by etching in buffered HF solution followed by annealing in oxygen. The growth process was monitored in situ with high pressure reflection high energy electron diffraction (RHEED). The RHEED intensity oscillation as shown in Fig. S1(a) were used to precisely control the number of atomic layers of SrIrO$_3$ and SrRuO$_3$. During growth of both SrIrO$_3$ and SrRuO$_3$, the substrate was maintained at 700 ℃ in an oxygen partial pressure of $10^{-1}$ mbar. The laser fluence and frequency were set to 1.4 J/cm$^2$ and 2 Hz, respectively.

**Electric and magnetic properties measurements.** Transport measurements were performed using a Physical Properties Measurement System (PPMS, Quantum Design). Resistivity and Hall measurements were carried out using the van der Pauw method on 5mm×5mm samples. The AHE was obtained by symmetrisation, which subtracts the magnetoresistance signal. Magnetic properties, including temperature-dependent magnetization and hysteresis loops, were measured using a superconducting quantum interference device (SQUID).

**ARPES measurements.** In-situ angle resolved photoemission spectroscopy (ARPES) measurements were performed on freshly deposited SrIrO$_3$/SrRuO$_3$ heterostructures in a PLD chamber directly connected to the ARPES chamber at the SIS beamline of the Swiss Light Source [46,47]. The photon energy was varied to locate the Γ plane of the electronic structure. The presented data presented were acquired at a temperature of approximately 20K and photon energy of 69eV.

## Acknowledgments

This work was supported by the National Key Research and Development Program of China (Nos. 2019YFA0307800, 2017YFA0303600), the National Natural Science Foundation of China (Nos. 12174406, 11874367, 51931011, 52127803), K.C.Wong Education Foundation (GJTD-2020-11), the Ningbo Key Scientific and Technological Project (Grant No. 2022Z094), the Ningbo Natural Science Foundation (No. 2023J411), the Natural Science Foundation of Zhejiang province of China (No. LR20A040001) and "Pioneer" and " Leading Goose" R&D Program of Zhejiang (2022C01032). M.R acknowledges the support of SNF Project No. 200021_182695.

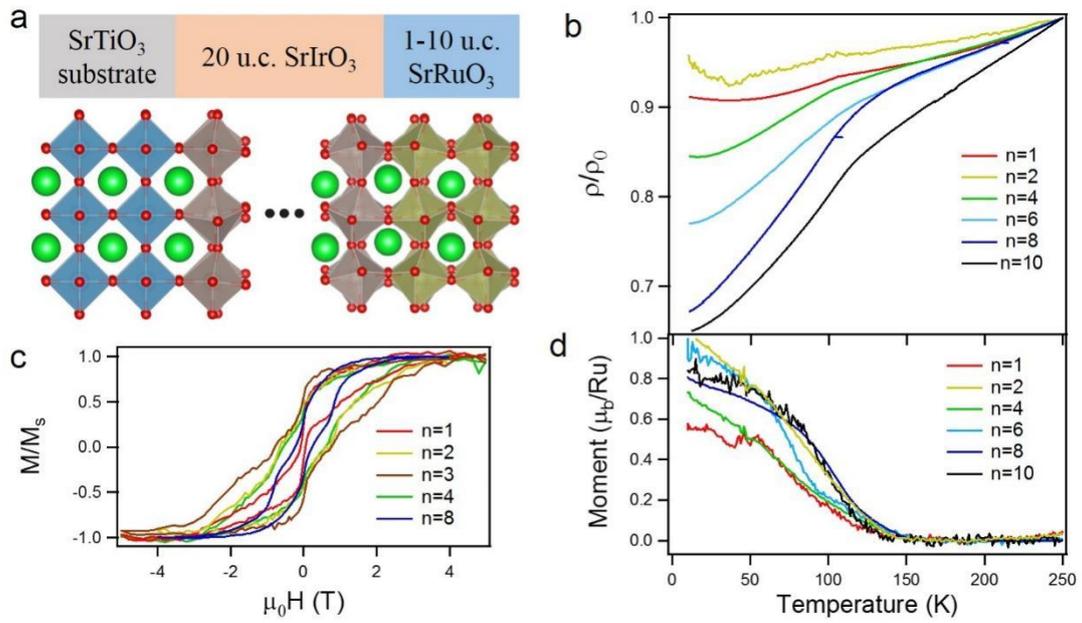

Figure 1 (a) Schematic illustration of the [SrIrO$_3$]20/[SrRuO$_3$]$n$ (20+$n$) heterostructure, consisting of a 20 u.c. SrIrO$_3$ film combined with an $n$ u.c. SrRuO$_3$ layer, grown on SrTiO$_3$ substrate. (b) Magnetic hysteresis loop of ultrathin SrRuO$_3$/SrIrO$_3$ heterostructures measured at a temperature of 10 K. (c) Temperature dependence of the electrical resistivity normalized to resistivity at 250K for the 20+$n$ heterostructures. (d) Temperature dependence of the magnetization for 20+$n$ heterostructures measured under a magnetic field of 1000 Oe.

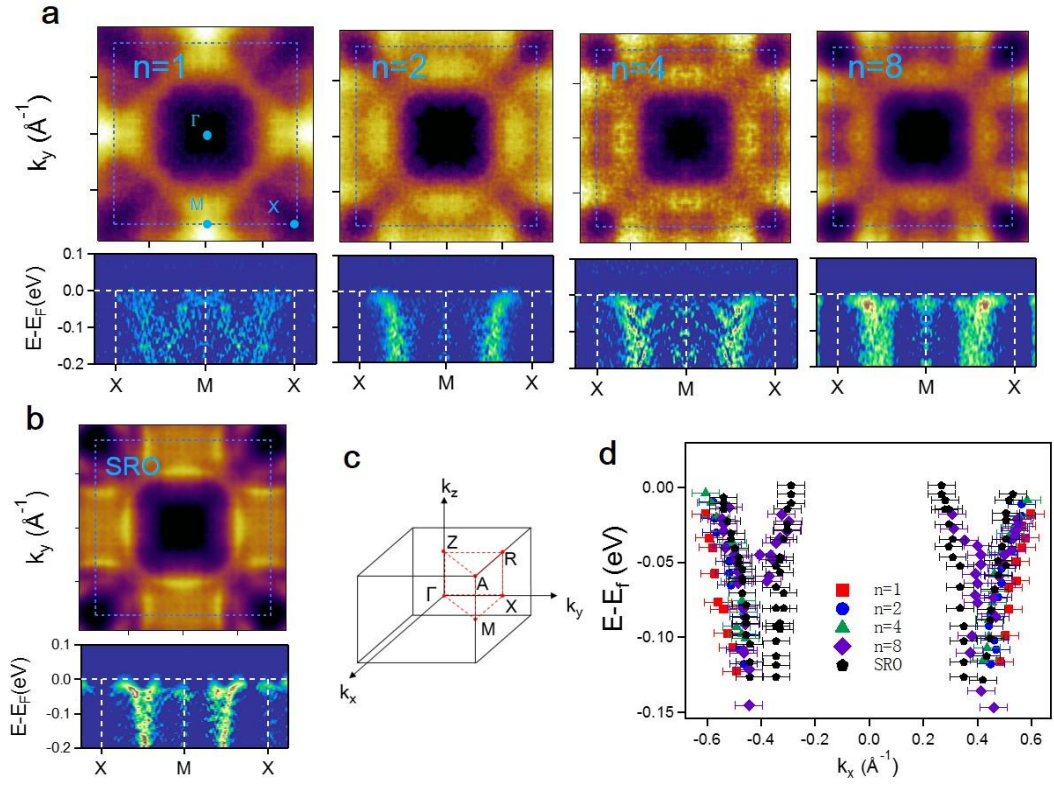

Figure 2 ARPES measurements [SrIrO$_3$]5/[SrRuO$_3$]$n$ ($n$=1,2,4,8) heterostructures and SrRuO$_3$ film. (a) Isoenergy intensity maps near the Fermi energy for heterostructures and second-derivative maps of the energy dispersion along the X-M-X direction. (b) Isoenergy intensity map near the Fermi energy and second-derivative map of the energy dispersion along the X-M-X direction for a SrRuO$_3$ film. (c) Schematic of the first Brilluion zone of SrRuO$_3$. (d) Fitted intensity maxima from the second derivative maps in (a) and (b) for different films.

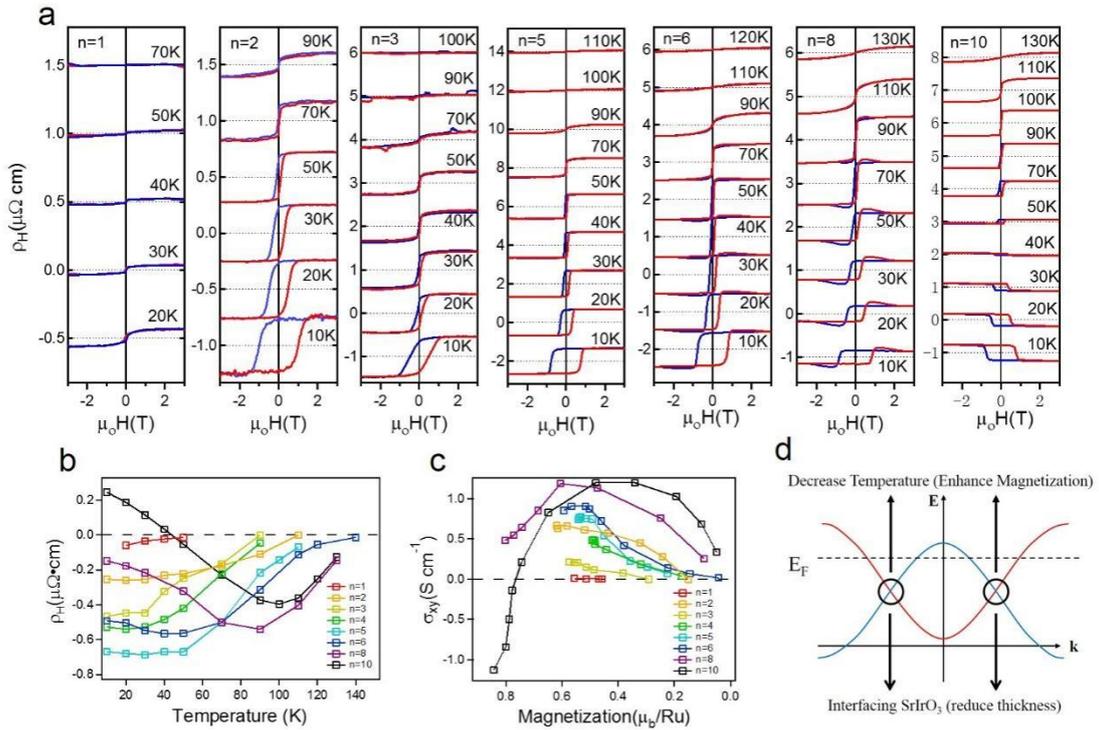

Figure 3 (a) Anomalous Hall effect measured at various temperatures for $SrIrO_3/SrRuO_3(20+n)$ heterostructures with different $SrRuO_3$ thickness ($n$=1,2,3,5,6,8,10). The normal Hall effect and the magnetoresistance signal are subtracted by performing symmetrisation. (b) Temperature dependence of the AHE signal is summarized for various heterostructures. (c) Hall conductance plotted against the magnetization obtained from the M-T measurements shown in Fig. 1(d). (d) Schematics illustrating the modification of the band structure by decreasing temperature and interfacing with $SrIrO_3$.

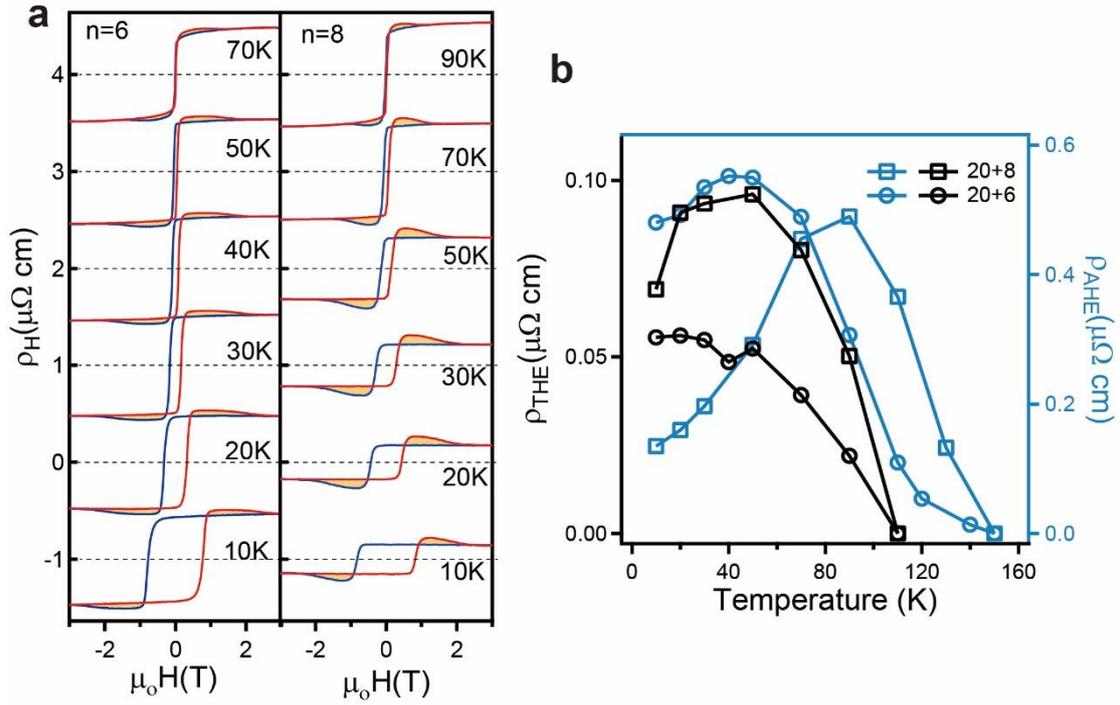

Figure 4 Topological Hall effect in SrIrO$_3$/SrRuO$_3$ heterostructures. (a) Hump-like feature observed in the AHE signal of 20+6 and 20+8 heterostructures. (b) Comparison of the AHE signal and the THE signal in 20+6 and 20+8 heterostructures, where the AHE sign remains unchange with temperature.